# An Approach to Secure Mobile Enterprise Architectures

Florian G. Furtmüller[1]

[1] **Solution Integration & Architecture Department, Computer Sciences Consulting Austria GmbH**
Vienna, AT-1200, Austria

**Abstract**
Due to increased security awareness of enterprises for mobile applications operating with sensitive or personal data as well as extended regulations form legislative (the principle of proportionality) various approaches, how to implement (extended) two-factor authentication, multi-factor authentication or virtual private network within enterprise mobile environments to ensure delivery of secure applications, have been developed.
Within mobile applications it will not be sufficient to rely on security measures of the individual components or interested parties, an overall concept of a security solution has to be established which requires the interaction of several technologies, standards and system components. These include the physical fuses on the device itself as well as on the network layer (such as integrated security components), security measures (such as employee agreements, contract clauses), insurance coverage, but also software technical protection at the application level (e.g. password protection, encryption, secure container).
The purpose of this paper is to summarize the challenges and practical successes, providing best practices to fulfill appropriate risk coverage of mobile applications. I present a use case, in order to proof the concept in actual work settings, and to demonstrate the adaptability of the approach.
**Keywords:** *Mobile Security Architecture, Authentication, Security, Integration Architecture & Interoperability, Privacy & Trust.*

## 1. Introduction

As a consequence of technological advances, it has become possible to fully integrate various actuator technologies as well as mobile devices into enterprise infrastructures and secure environments. These developments open up a huge amount of innovative interaction scenarios, involving new forms of user communication and behavior. Therefore, enabling companies to manage policy, security and support of mobile devices became a major issue. Enterprise mobile device management (MDM) software is evolving to offer cross-platform device support, vendor independence and keeping the focus on a secure integration layer [17].

This is an opportunity for exploring the potentials and perspectives of mobile enablement of enterprises supporting collaborative work that enables employees to increase their productivity (up to 20 % [20]) and efficiency, as well as flexibility and accessibility. Faster networks and extended battery life offer the support of a hand-held microcomputer as personal assistant. This hand held device collects an enormous amount of personal data, ranging from the user's email address to location, contact list, calendar & photos and tether it to a single unique device ID number [4].

Motivated by these developments, its close integration and interaction between business and private usage this paper can be seen as reference and aims bringing together various challenges (refer to section VI):

--Various data (private or business data)
--Access points (internet, intranet, public Wi-Fi…)
--Sensitive device data (daily financial sales reports)
--Application content (business or private applications)
--Corporate data access (secure or non-secure)
--Malware / Phishing software
--Social media usage
--Bring-your-own-device (BYOD)
--Online & offline availability
--News headlines (malpractice).

Clustering the challenges above will lead to the following security risk domains, physical risk (lost or stolen device), access risk (login or network accessibility of not authenticated persons e.g. man-in-the-middle), usage risk (software bugs, jail-breaking, malware) and memory risk (private or sensitive data stored on device (or even worse stored on removable SD card) [20].

Meeting all the multiple challenges and risks in which a mobile enabled enterprise is confronted with must be discussed and clearly clarified before defining an enterprise mobile security strategy. Providing software for mobile devices, calls for context-, business- and data-dependent analysis already at design phase and requires a framework to manage IT architecture (EAM) [14].

The basic concepts of authentication processes, for two-factor authentication and multi-factor authentication, as well as for a virtual private network approach (see III.B) are used to meet the requirements given above.

I have identified two main security concepts, which meet the requirements, mentioned in I and II, appropriately. Section III presents the concepts that were applied when designing a communication channel of designated applications. In section III and IV, implementation of the approach and application during design time and in operation are described. Section V shows how the concept has been used in practice. Finally, I conclude with a summary of achieved results and some inspirations for future work.

## 2. Requirements

The lack of time in development phase and the pressure for releasing new applications - especially in the mobile applications market - are the main reasons why developers neglect security issues. Besides automated code analysis software and sophisticated test procedures, it is recommended to perform security testing during the quality assurance (QA) phase and predefine already approved standard libraries for development.

However, the mentioned approaches in section III, TFA, MFA and VPN in combination with adequate MDM solution allow clarification of problem areas, such as protecting corporate data in transit over public Wi-Fi or cellular networks, encrypting data stored on device, disabling device communication modules and hardware features, authenticating device or user using certificates or domain security credentials, malware protection and intrusion detection, preventing harmful internet downloads and unauthorized software installation or strategies for dealing with lost or stolen devices [23] already at design time.

## 3. Approach

For the development of our best practices, I have reviewed earlier approaches in terms of concepts and implementation strategies. The concepts, that have influenced the presented approach, are described in the following section:

3.1 Two-Factor Authentication (TFA, MFA)

Two-factor authentication represents the combination of knowledge with possession of an object.

*Concept*
This approach combines the knowledge of a secret (e.g. password) with the possession of a clearly identifiable object (e.g. token, a special USB key or smart card) or personal characteristics (e.g. biometrics). Extended TFA or multi-factor authentication combines more methods [5].

*Example:* Maestro debit card, logical knowledge (PIN) and physical object (map).

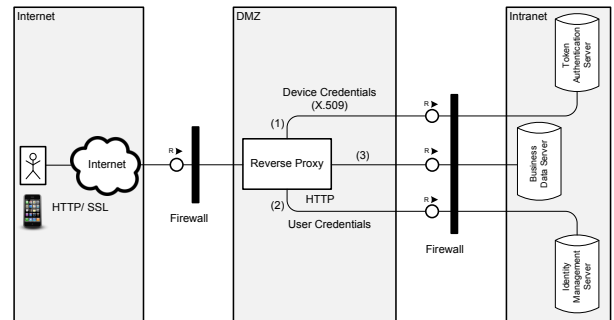

Fig. 1. Two-Factor Authentication (TFA, MFA) Process (refer [4])

*Process*
As shown in Fig. 1, implementation stack of a two-factor authentication follows the process, (1) authentication of the object (for example via the mobile device installed X.509 certificate), (2) authentication of the property (e.g. password) and as positive result establishes (3) connection to business server. This will lead to strong authentication that meets compliance requirements, provides improved security and minimized risk through a $2^{nd}$ factor as well as eliminating phishing with one-time passwords and minimized 'window-of-opportunity' but does not prevent the possibility of 'man-in-the-middle' attacks. Therefore, it will be necessary to establish complex passwords or layers (that will lead to increased training and operational costs) and a contrary waiver of complex passwords requires additional security hardware [19].

3.2 Mobile Virtual Private Network (Mobile VPN)

As already established in various enterprises virtual private networks (VPNs) extends the combination of knowledge with possession of an object.

*Concept*
In addition to the PIN of the user a second security code (via VPN token) will be required during authentication. This allows establishing a tap- and tamper-proof VPN tunnel as well as encrypting network packets [6].

*Example:* Home office via VPN, logical knowledge (PIN) and a physical object (SecureID token for One-Time-Passwords)

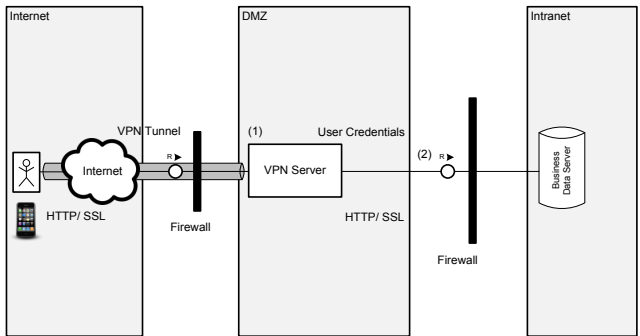

Fig. 2. Virtual Private Network Process (refer [6])

*Process*

Fig. 2 depicts the process flow of establishing a VPN connection, starting with (1) authentication and setting up a VPN tunnel to the VPN server (e.g. via RSA Token + PIN). After successful authentication the end-user will be (2) connected to the business server (e.g. SAP ERP backend). VPN incorporates a secure network connection with the possibility of encrypted data packets, scalability (ease of adding or removing users) and standardized by virtual private network consortium (VPNC).

This approach will require additional hardware (e.g. for an RSA token) and encourages users to enter password credentials and token. On the contrary Quality-of-Service (QoS) relies on Internet Service Provider (ISP) which has direct impact on service stability and performance.

More information regarding to user authentication solutions can be found here [26].

## 4. Implementation & Application

As already defined in the requirements (see, section III and IV), mobile applications are designed to work in secured and non-secured environments. To provide the basic service infrastructure, a mobile device management (MDM) solution should be evaluated upon context and enterprise specific strategic requirements. This kind of middleware embedded in a web-oriented architecture (WOA) provides simple and transparent management of mobile devices over various environments and heterogeneous platforms and features device procurement, asset inventory management, reporting, logging, policy enforcement and management, help desk support, remote configuration and assistance to name but a few.

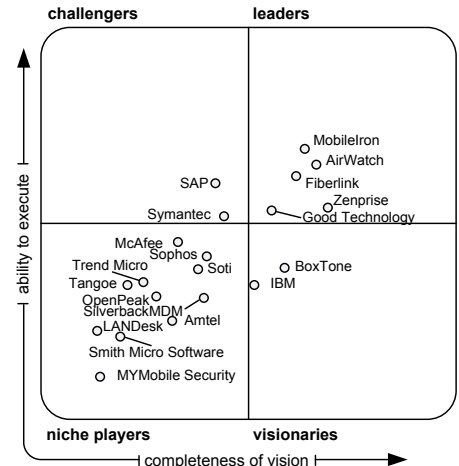

Fig. 3. Magic Gartner Quadrant Mobile Device Management (see [18])

Hence, to customer technical requirements and specific feature requests (see V) I had to evaluate several MDM solutions such as AirWatch, BoxTone MDM, Fiberlink's Maas360 (as pure SaaS), Good Technology's Mobile Control, Matrix42, Enterprise Mobility Management from McAfee, Symantec's Mobile Management and Zenprise an overview of established MDM solutions can be found in Fig. 3, for detailed information regarding to critical capabilities for MDM solutions refer to [17].

To be prepared for upcoming future enterprise requirements and to ensure a viable mobile security infrastructure, a careful selection of corporate MDM solution for implementing mobile security concepts is essential.

### 4.1 Use of Design Patterns and Frameworks

Object-oriented design patterns (refer to [7]) are commonly accepted means to construct highly structured software that is easy to grasp. Hence, upcoming mobile applications should be developed using established patterns. In this way, existing applications can easily be extended, and its components can easily be modified and exchanged with others (significant reduction of maintenance costs [7]). As an example for the concrete use of frameworks for extended TFA (MFA), Fig. 4 depicts the process flow provided by [29]. To add further factor to the authentication procedure could be realized by simply implementing an extension e.g. dynamic generated password for user login provided by SMS gateway (see Fig. 5).

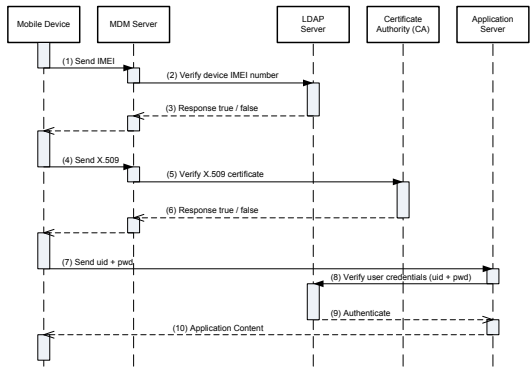

Fig. 4. Process Simulation - Extended Security via Static Password

Fig. 4 shows how the authentication process operates, as first factor, device IMEI number (unique number to identify mobile devices) is sent to the MDM solution to verify if device belongs to the group of enterprise managed devices or not. As second factor, a X.509 certificate stored in the web container on the device (see, Fig. 6) will be requested and verified. Finally, as an extended factor, a combination of username and password has to be transmitted to the back-end application server (e.g. SAP). As of this moment, users are authenticated and the back-end service is delivering requested data.

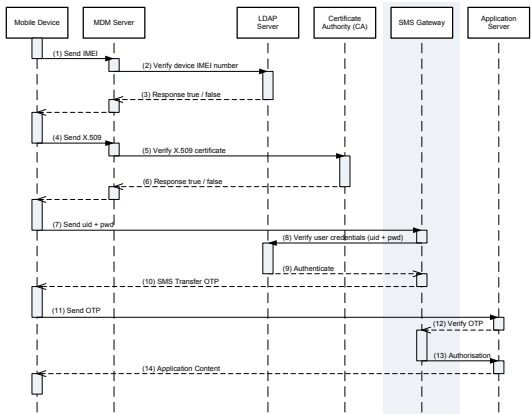

Fig. 5. Process Simulation - Extended Security via One-Time-Password

A reduction of the window-of-opportunity can be achieved by setting up a SMS gateway for issuing One-Time-Passwords (OTP) or Time-based One-Time Passwords (TOTP), as illustrated in Fig. 5. An OTP reference implementation can be found here [15]. Additionally, the HOTP (HMAC-based One Time Password) algorithm represents an open standard for event-based OTP authentication and its community edition is freely available [25].

Alternatives, such as WiKID Strong Authentication System [29] a dual-source, software-based (APIs available for PHP, Java, Ruby, Python and C#) two-factor authentication system (applicable via VPN as well), its flexible design impress with more extensibility than hardware token based solutions.

A popular representative of TFA via TOTP [11] is the open source application *Google Authenticator* available for all Gmail and Google Apps users, details and source can be found on the project page [9].

### 4.2 Related Work

For the development of our mobile security concepts, I have reviewed earlier approaches in terms of concepts and implementation strategies. The projects - that have influenced the presented approach - are described below:

*Project: FACTOR TWO*
This approach is built with Sencha Touch mobile framework (see [21]) and provides a simple two-factor authentication mechanism based on SHA256 JS implementation from the *crypto-js* project [10]. FACTOR TWO has been setup for a particular user (via simple PIN exchange between website and mobile application) and a one-time token will be required for each login. If the mobile application has an internet access, the token could be submitted automatically by pushing a single button. Once a valid token had been submitted, the user is able to login into the site with his credentials. The token remains valid for a predefined time interval since it has been submitted (minimize window of opportunity, see section III). If user authentication fails during this time interval, the user will be forced to re-submit a new token. The service utilizes an offline caching mechanism that allows it to run without internet connection. In this mode the application issues a valid token that the users have to enter into login form along with their credentials [24].

*Project: QuickSec™ Mobile VPN Client*
This project keeps the focus on implementing an application with a specific VPN client, in the way that its VPN connection once established should be only available to the designated application and all other applications, which are installed on the mobile device should use common ISP internet connection.

Android Gingerbread [28] and recent versions, as well as iOS are including a VPN client. Android supports L2TP/IPsec (PPTP) protocols and iOS, SSL VPN, Cisco IPSec, L2TP over IPSec and PPTP. However, split-tunneling is also supported by Android [27] and iOS, this technology ensures that clients are supplied by the existing VPN connection and prevent initiatives to detour data from corporate network [22].

The split-tunneling concept for personal VPN allows users to access a public network commonly the Internet, at the same time that the user is allowed to access resources on the VPN. It avoids bottlenecks and cover bandwidth as the traffic to the Internet does not pass through the VPN server. A drawback of this method is that it essentially renders the VPN vulnerable to attack as it is accessible through the public, non-secure network [1]. This is the case of split-tunneling mechanism of Cisco client Juniper Networks SA2500 SSL VPN appliance. It offers a feature that allows users to work through split tunnel and enable them to route the traffic through different channels. Subsequently, QuickSec Mobile VPN Client offers complete IPsec support, IPv6, IKEv2 with MOBIKE, xAuth, EAP-based authentication and split-tunneling support (ibid.).

However, within this context SSL VPN connection is the most reliable one since features strong encryption and security [16].

In addition to Cisco or QuickSec the following solutions for establishing reliable connections for mobile devices do exist: IAPS Security Services VPN, Hide My Ass! VPN and PureVPN.

## 5. Proof of Concept

The correct operability and stability of the approach was first tested using simulated components. The practical feasibility was then shown in a real world scenario with an invoice approval process.

5.1 Process Simulation

Hence I developed a process that achieves most of the requirements, mentioned in [I and II]. Therefore I created the following test infrastructure, containing an MDM solution [17] for the central administration of mobile devices (IMEI number will be mapped to user account e.g. LDAP or active directory by the administrator). In a next step, the designated user will receive a download link for the new service or download the application directly from the private application store. Subsequently, during the installation of the app the users will be asked to download and install their personalized X.509 certificate, this X.509 certificate will be stored in the Web Container of the app and cannot be modified or copied by the users.

On the contrary, the Web Container will be administrated by MDM solution (e.g. application updates, wipe-out of stolen or lost devices). As a result the user will now be able to connect to the back-end server. This scenario realized using the multi-factor authentication (MFA) concept (see, section III.A).

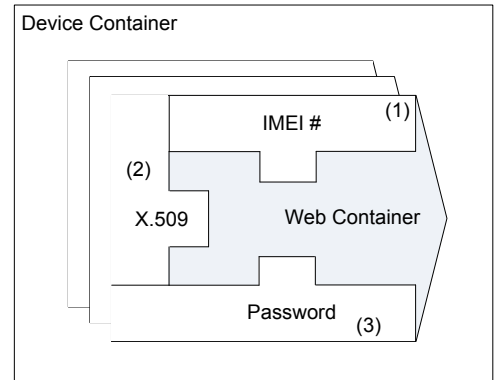

Fig. 6. Simulation – Multi-Factor Authentication (MFA, Extended TFA)

First (1), users must provide device number (IMEI#). As a next step the private user certificate (2) has to be transmitted and verified and finally users must transfer valid logon credentials (3). The logon credentials must be entered by the end user each login time; X.509 (web container) and IMEI# (device container) are stored on the mobile device and will be provided for authentication automatically.

Fig. 6. depicts the following implementation:

1) Device Management (IMEI#) check with registered corporate mobile devices
2) X.509 Certificate User stored by user on mobile device (once a time) check with employee and IMEI#
3) a- Static Password (defined by user) check password e.g. MD5 encrypted
4) b- Dynamic Password (request 'one-time-password' (OTP), generated password sent by SMS) e.g. validity 5 min (TOTP).

5.2 An Actual Scenario of Use

Here, an invoice approval process was chosen for demonstration. This process consists of three actors: requestor of the invoice approval (commonly known as accounting assistant) and two approvers (approvers are two signing authorities). The accounting assistant is - among other things - responsible to create, submit, manage and clone requests.

Therefore, it was required to set up a separate security level for the accounting assistant as well as an extended security level for the authorized signatories. Once the invoices have been approved, they will be sent to Accounting for verification and payment (see Fig. 7).

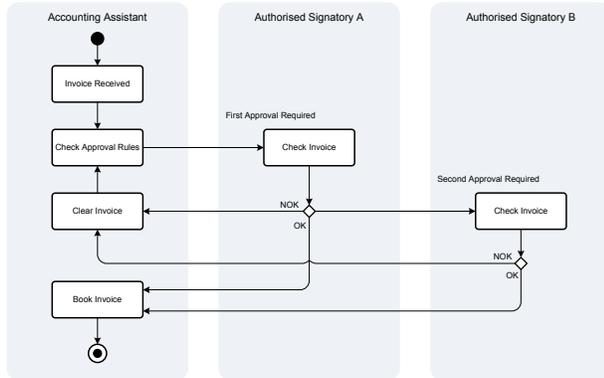

Fig. 7. Scenario of Use - Invoice Approval Process

As a next step it was essential to map approval rules to the enterprise security levels. Therefore, invoices below a minimum limit of € 500.-, recurring invoices or invoices authorized by a contract to which the accounting staff has access do not require approval and can be signed off by the accounting assistant. The amount limit of invoices which can be approved without two authorized signatories are defined with € 100,000.-. To approve an invoice within this range a medium security level was implemented. All invoices above this value must be approved by two authorized approvers, which require a separate security level too - defined as high security level.

Security level descriptions, transformations and technical requirements,

   --Low security level -> manage and forward requests, payment sign off for invoices below € 500.- (by accounting assistant) requires IMEI#, X.509 and password (see Fig. 6).
   --Medium security level -> one signing authority verification of invoice, approval and payment (above € 500.- to € 100,000.-) requires IMEI#, X.509 and password plus indexed TAN (acting as OTP via SMS Gateway).
   --High security level -> two signing authorities verification of invoice, approval and payment (above € 100,000.-) requires IMEI#, X.509 and password plus indexed TAN (acting as OTP via SMS Gateway).

Besides the test for correct operation, a mobile enterprise environment has to be set up. Therefore Fiberlink MaaS360 MDM solution [2] was chosen which allows issuing and verifying X.509 certificates (CA). As a next step, a connection to the corporate LDAP server for provisioning of user objects and user attributes, mapping of IMEI# and UID was established. To achieve a predefined medium and high security level, an OTP [25] service infrastructure (OTP Validation Server and SMS Gateway) [3], which is communicating directly with the mobile application, was implemented (see Fig. 5 & Fig. 8).

The mobile application for invoice approval was built with Sencha Touch [21], a cross-platform HTML5 Mobile application framework, which is based on JavaScript and allows deployment on various mobile devices (tested on Samsung Galaxy, Samsung Galaxy Tab 10, Apple iPhone, Apple iPad 2 and Blackberry Torch) without recompilation.

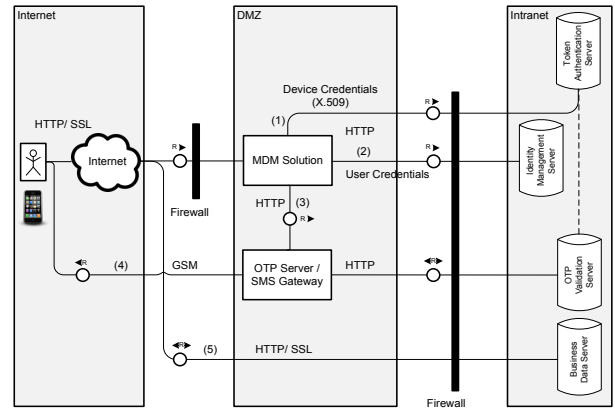

Fig. 8. Solution Architecture - Mobile Invoice Approval Architecture

The integrated environment proved to be successfully orchestrated by the given configuration and allows real-time, asynchronous approving of invoices. In the next stage of the evaluation, I aim at the dynamic extension of the tracked workbench area by attaching further authentication factors (refer to III.A) when necessary.

## 6. Conclusions

At the very beginning of mobile projects it is recommended to perform a situation analysis, to clarify what is already established in the company, how it can be used, what needs to be procured - how the enterprise can be mobile enabled. It is essential to clearly define requirements (e.g. data quality, integration, single-sign-on, performance, functional and non-functional requirements), perform cost-benefit analysis, verify possible advantages and disadvantages of the planned solution consider legal regulations (e.g. data loss on wipe out of mobile device) and follow the trend.

$$Project\ Duration = \frac{Effort}{Number\ of\ Persons} \qquad (1)$$

Consider Eq. (1), it is commonly accepted but the relation between *'Effort'* and *'Number of Persons'* cannot be seen as valid within software projects and even not for mobile software projects. For this reason it is very important to start a project with prototyping or a simple proof of concept, keep the team small and clearly clarify customer expectations and requirements. In most of the cases mobile projects consists of various infrastructure layers and heterogeneous technologies or components, this will automatically lead to high complexity including steep learning curve and high effort in coordination, communication of all involved parties. 'If you do not actively attack the risks in your project, they will actively attack you' [8].

The presented approach serves to establish several safety standards, especially for mobile business applications within enterprise environments.

All identified requirements could be met both, from a conceptual perspective, and at the implementation level. I was also able to demonstrate user benefits that are as follows: sustainable technology, state-of-the-art security concepts, consideration of legal regulations, combined with years of experience and proven best practices. Additionally, following the presented approaches will minimize implementation risk as well as it will lead to reduce time-to-market of mobile projects.

The concepts mentioned above bring together related topics from security, project management, technology and legal, thus can be seen as cross-divisional procedure or methodology for implementing secured mobile business applications.

|  | Low Embeddedness | High Embeddedness |
|---|---|---|
| **High Mobility** | Mobile Computing | Ubiquitous Computing |
| **Low Mobility** | Traditional Computing | Pervasive Computing |

Fig. 9. Future Prospective - Mobile Computing, see [13]

Traditional computing is becoming less important in the future. In contrast, the market is moving increasingly towards ubiquitous computing through to pervasive and mobile computing (see Fig. 9). It is foreseeable that user interfaces will be pushed into the spotlight even more and will provide access to computers for a wide audience (refer to [13]).

In future projects, I will evaluate the approaches technology base and its applicability with VPN split-tunneling in combination with dynamic passwords as Time-based One-Time-Passwords (TOTP) transmitted on SMS gateway infrastructure.
Placing a reverse HTTP proxy in front of the MDM solution that verifies signed OAuth (Open Authorization protocol IETF) requests from mobile devices could also be an option, the claim for universal applicability has still to be validated.

"If you don't know where you are going, any road will take you there." *Lewis Carroll*

## Acknowledgments

I would like to thank Jürgen Mairhofer (Department of Biotechnology, University of Natural Resources and Life Sciences, Vienna, Austria), Dietmar Moltner (Computer Sciences Consulting Austria) and Michael Kaufmann (Computer Sciences Consulting Austria) for their input and constructive criticism.

## References

[1] An Introduction to IP Security (IPSec) Encryption, Cisco System, 2012.
http://www.cisco.com/en/US/tech/tk583/tk372/technologies_tech_note09186a0080094203.shtml (June 2012)

[2] Diodati M., Mobile Device Certificate Enrollment: Are You Vulnerable?, Gartner Blog, 2012.
http://blogs.gartner.com/mark-diodati/2012/07/02/mobile-device-certificate-enrollment-are-you-vulnerable (July 2012)

[3] Diodati M., The Evolving Intersection of Mobile Computing and Authentication, Gartner, 2011.
http://www.gartner.com/id=1882514 (July 2012)

[4] Etherington D.: Sleepless? Then Stop Taking Your iPhone To Bed, Giga Omni Media, Inc., 2011.
http://gigaom.com/2011/05/23/report-mobile-workers-in-bed-with-smartphones (June 2012)

[5] Fadi Aloul, Syed Zahidi, Wassim El-Hajj, Two Factor Authentication Using Mobile Phones, IEEE/ACS 2009.

[6] Ferguson P., Huston G., What is a VPN?, Cisco Systems, 1998.
http://citeseerx.ist.psu.edu/viewdoc/summary?doi=10.1.1.28.972 (July 2012)

[7] Gamma E., Helm R., Johnson R., Vlissides J.: Entwurfsmuster, Elemente wiederverwendbarer objektorientierter Software, Addison-Wesley Verlag, 2004.

[8] Gilb T., Finzi S.: Principles of software engineering management, Addison-Wesley Pub. Co., 1988.

[9] Google Open Source Project, Google Authenticator (TOTP), Google Inc., 2011.
http://code.google.com/p/google-authenticator/ (June 2012)

[10] Google Project, JavaScript implementations of standard and secure cryptographic algorithms, Google Inc., 2012
http://code.google.com/p/crypto-js/ (June 2012)

[11] Internet Engineering Task Force, TOTP Time extension of HMAC-based One-Time Password algorithm, 2011.
http://www.rfc-editor.org/rfc/rfc6238.txt (June 2012)

[12] Lehner F.: Mobile und drahtlose Informationssysteme: Technologien, Anwendungen, Märkte, Springer, 2003.

[13] Lyytinen K., Yoo, Y.: Issues and Challenges in Ubiquitous Computing, Communications of the ACM, 2002.

[14] Buckow H., Rey S. - Business Needs IT Architecture, McKinsey & Company, 2010.

http://www.mckinseyquarterly.com/Why_business_needs_should_shape_IT_architecture_2563 (June 2012)

[15] OATH Initiative, HOTP Algorithm - One-Time-Password Reference Implementation, 2004. http://rfc-ref.org/RFC-TEXTS/4226/chapter16.html (June 2012)

[16] QuickSec™ Mobile VPN Client for Android, AuthenTec, Inc., 2011. http://www.authentec.com/Products/EmbeddedSecurity/SecurityToolkits/QuickSecVPNAndroid.aspx (June 2012)

[17] Redman P., Basso M., Critical Capabilities for Mobile Device Management, Gartner Inc., 2011. http://www.gartner.com/DisplayDocument?doc_cd=213877 (June 2012)

[18] Redman P., Girard J., Wallin L. O., Magic Quadrant for Mobile Device Management Software, Gartner Inc., 2012. http://www.gartner.com/DisplayDocument?doc_cd=211101 (June 2012)

[19] Schneider B.: Two-Factor Authentication: Too Little - Too Late, Communications of the ACM, 2005.

[20] Schurek A.,Das mobile Industrieunternehmen, IBM, 2012. http://www-05.ibm.com/de/events/swg-solutions/pdf/Das-mobile-Industrieunternehmen-external_2012-05-22.pdf (June 2012)

[21] Sencha Touch 2, A high-performance HTML5 mobile application framework, Motorola Mobility Inc., 2012. http://www.sencha.com/products/touch (June 2012)

[22] Shinder T., Remote Access VPN and a Twist on the Dangers of Split Tunneling, TechGenix Ltd., 2005. http://www.isaserver.org/tutorials/2004fixipsectunnel.html (June 2012)

[23] Symantec Security Analysis of iOS and Android, Symantec Corp., 2011. http://www.symantec.com/about/news/release/article.jsp?prid=20110627_02 (June 2012)

[24] Tchijov A., Project FACTOR TWO: Factor-2 based Authentication built on Sencha Touch Framework), 2011. http://drupal.org/project/factortwo (June 2012)

[25] The Internet Society, Reference HOTP Algorithm - An HMAC-Based One-Time Password Algorithm, 2005. http://www.ietf.org/rfc/rfc4226.txt (June 2012)

[26] VeriSign Benutzerauthentifizierungslösungen, Symantec Corp., 2012. http://www.symantec.com/de/de/user-authentication (June 2012)

[27] Marg C., VPN Konfiguration - VPN unter Android nutzen, TU Clausthal, 2012. https://doku.tu-clausthal.de/doku.php?id=vpn:vpn_unter_android_nutzen (June 2012)

[28] The Android Open Source - Android VPN Profile Reference Implementation, 2009. http://www.java2s.com/Open-Source/Android/android-core/platform-frameworks-base/android/net/vpn/VpnProfile.java.htm (June 2012)

[29] WiKID Strong Authentication System, Open Source Two-Factor Authentication, 2011. http://www.wikidsystems.com/community-version (June 2012)

**Florian G. Furtmüller** studied business informatics at the Johannes Kepler University of Linz, Austria. He earned his master degree in 2007 and subsequently, started his work as Enterprise Portal Consultant for SAP and IBM systems at IDS Scheer Austria GmbH.

He has more than five years of experience in application integration architectures SOA, WOA, EDA, MDA and design patterns, especially with enterprise portals and components. He is currently working for Computer Sciences Consulting Austria GmbH, department for Solution Integration & Architecture as Senior Consultant and Application Architect.

Mr. Furtmüller is co-author of A Tuple-Space based Middleware for Collaborative Tangible User Interfaces in Proceedings of WETICE 07, IEEE Press, 2007, ISBN 0-7695-2879-1, a collaboration with Stefan Oppl.